\documentclass[reprint,amsmath,amssymb,showpacs,aps,prl]{revtex4-1}
\usepackage[utf8]{inputenc}
\usepackage[T1]{fontenc}
\usepackage[english]{babel}
\usepackage{graphicx}               % Include figure files
\usepackage{placeins}             % Prevents figures to float further than \FloatBarrier
\usepackage{hyperref}               % add hypertext capabilities
\usepackage{color}                  % Allows to use colors in the text
\usepackage{csquotes}               % Easy handling of quotation marks
\usepackage{nicefrac}               % Allows to generate fractions like a/b

\begin{document}

%%%%%%%%%%%%%%%%%%%%%%%%%%%%%%%%%%%%%%%%%%%%%
%%%%%       Author Information          %%%%%
%%%%%%%%%%%%%%%%%%%%%%%%%%%%%%%%%%%%%%%%%%%%%

%\preprint{APS/123-QED}

\title{Radially Self-Accelerating Beams}

\author{Christian Vetter}
%\email{Christian.Vetter@uni-jena.de}
%\altaffiliation[Also at ]{Physics Department, XYZ University.}
\author{Toni Eichelkraut}
\author{Marco Ornigotti}
\author{Markus Gräfe}
\author{Alexander Szameit}%
\email{Alexander.Szameit@uni-jena.de}
\affiliation{Institute of Applied Physics, Friedrich-Schiller-Universit\"at Jena, Germany}%

%\affiliation{Second institution and/or address}
%\affiliation{Third institution, the second for Charlie Author}
%\affiliation{Authors' institution and/or address}

\date{\today}

%%%%%%%%%%%%%%%%%%%%%%%%%%%%%%%%%%%%%%%%%%%%%
%%%%%               Abstract            %%%%%
%%%%%%%%%%%%%%%%%%%%%%%%%%%%%%%%%%%%%%%%%%%%%

\begin{abstract}
We report on optical non-paraxial beams that exhibit a self-accelerating behavior in radial direction. Our theory shows that those beams are
solutions to the full scalar Helmholtz equation and that they continuously evolve on spiraling trajectories. We provide a detailed insight into the
theoretical origin of the beams and verify our findings on an experimental basis.
\end{abstract}

\pacs{42.25.Fx, 42.25.Hz, 42.60.-v}

%%%%%%%%%%%%%%%%%%%%%%%%%%%%%%%%%%%%%%%%%%%%%
%%%%%           Introducation           %%%%%
%%%%%%%%%%%%%%%%%%%%%%%%%%%%%%%%%%%%%%%%%%%%%

\maketitle
%\section{\label{sec:intro}Introduction}
\FloatBarrier Self-accelerating wave packets freely accelerate even without any external potential present. This intriguing phenomenon is of rapidly
growing interest since its advent in optics in 2007 \cite{Siv07a, Siv07, Ban08, Min11, Zha13, Kam13}. The most prominent example of self-accelerating
waves has been introduced by Siviloglou and coworkers \cite{Siv07, Siv07a}: They demonstrated, that an Airy-type wave packet exhibits a linear
transversal acceleration and is therefore following a parabolic trajectory.

Enlarging the scope and the versatility of Airy beams, nonparaxial generalizations in terms of full vectorial solutions of Maxwell's equations \cite{Kam12} as well as by the method of caustics \cite{Cou12} were investigated.
%
%OLD VERSION
%they were later generalized to the non-paraxial case by solving the full vectorial Maxwell equations \cite{Kam12} as well as by the method of caustics \cite{Cou12}. 
%
Their curved trajectories render classical Airy beams a powerful tool in many areas of application. For instance, in the field of particle manipulation, micro beads have been
guided and sorted in a new fashion \cite{Bau08} beyond the scope of classical optical tweezers.  Moreover, it was shown that curved plasma channels
have many advantages over their straight counterparts, e.g., when it comes to the spatially resolved detection of secondary signals \cite{Pol09}.
Additionally, Airy wave packets inspired excessive fundamental research in the field of nonlinear optics \cite{Kam11,Lot11,Dol12,Bek11}, and boosted
the study of waves with intensity maxima that propagate along almost arbitrary trajectories \cite{Gre11}. Broadening the range of influence beyond
the scope of optics, Airy beams have been utilized in electron beam shaping \cite{Vol13} as well.

\begin{figure}[b]
\includegraphics[width=\linewidth, page=1]{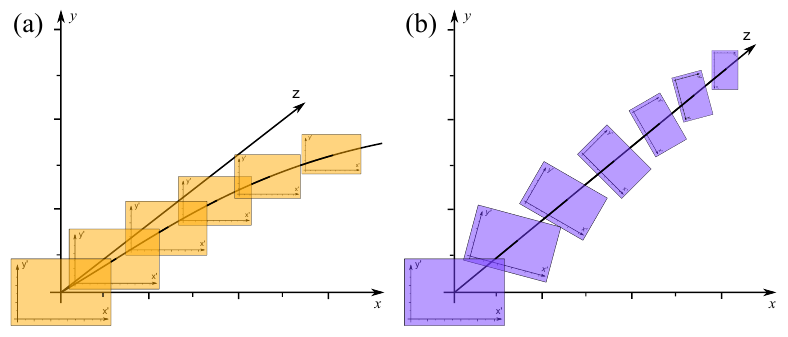}
\caption{Illustrative presentation showing the accelerative behavior of Airy beams (a) and radially self-accelerating field\,/\,intensity distributions (b).}
\label{fig:fig1}
\end{figure}

One common feature of all of the aforementioned waves - even in two-dimensional settings - is that they accelerate linearly, namely  along a
specific Cartesian coordinate axis (see Fig.~\ref{fig:fig1}~(a)). This obvious limitation brings about a number of fundamental questions: Is it
possible to generate optical beams that show a self-accelerating behavior along different types of trajectories? Could such wave packets be
shape-invariant or even non-diffractive? Based on the numerous already existing applications of Airy-type beams discussed in the previous paragraph,
it should be obvious that beams for which the aforementioned questions
%
%OLD VERSION
%the preceding questions 
%
can be answered affirmatively would enrich the optical toolbox in many areas of
application and research.  Moreover, from a fundamental point of view, it is essential to determine under which kind of approximations analytical
solutions for such beams can be found. In other words, will those solutions be restricted to the paraxial case or do they obey the scalar Helmholtz equation
or even Maxwell's equations?

In the present article, we report on a new class of self-accelerating diffraction-free waves that move along three-dimensional spiraling
trajectories. As such, they behave as if they were influenced by a radially symmetric external potential even though the propagation takes place in
free space. Observed from a rotating, co-moving frame of reference like the one depicted in Fig.~\ref{fig:fig1}~(b), the beams we present are
propagation-invariant. Within a thorough theoretical discourse we will derive a general, explicit, and analytical expression for this new class of
beams. We will show the great versatility of these beams, for which - while outperforming Airy-type beams - the transverse cross-section is highly
tunable after fixing predefined spiraling properties. In addition, we are going to verify our theoretical findings on an experimental basis utilizing
an intuitive understanding borrowed from Fourier-optics. In this regard, we present a simple yet powerful setup that enables one to address the
entire parameter range of the presented beams.

%%%%%%%%%%%%%%%%%%%%%%%%%%%%%%%%%%%%%%%%%%%%%
%%%%%               Theory              %%%%%
%%%%%%%%%%%%%%%%%%%%%%%%%%%%%%%%%%%%%%%%%%%%%

%\section{\label{sec:theory} Theory}
Our theoretical section is comprised of two parts.  The first one consists of an extensive theoretical derivation regarding the most general expression of a beam which exhibits a field pattern that is invariant in a rotating frame of reference. As it will be discussed later, of more practical interest might be the implementation of beams with a rotating intensity distribution. For this reason, in the second part of our theory section, we pose conditions on the beam intensity only (i.e., no condition is posed on the phases) finding a more general class of beams.

%\subsection{Rotating Field Distribution}
First, we want to model a beam that is propagation-invariant in a co-moving, rotating frame. This wave is supposed to be a solution to the scalar Helmholtz
equation $\Delta E +k^2 E=0$, where $E$ is a scalar electric field and $k=\nicefrac{2\pi}{\lambda}$ the corresponding wave number. Since we are
dealing with rotating solutions, it is a natural choice to work with cylindrical coordinates. Then, the most general solution of the scalar Helmholtz
equation can be written as \begin{align} E(r,\varphi,z)=\sum_{n=-\infty}^{\infty}\int_0^{\infty}d\alpha C_n(\alpha)J_n(\alpha r)e^{i(n\varphi+\beta
z)}, \label{eq:generalSuperpos}
\end{align}
which is essentially a superposition of fundamental eigenmodes given in terms of diffraction-free Bessel waves. The spatial structure of each eigenmode is determined by $J_n(\alpha r)e^{i(n\varphi+\beta z)}$, where $J_n(\alpha r)$ represents the Bessel function of order $n$ and
$\beta=\sqrt{k^2-\alpha^2}$ is the longitudinal component of the wave vector, or propagation constant. For an arbitrary beam, the expansion coefficients $C_n(\alpha)$ are arbitrary as well. In the following, we will derive conditions for $C_n(\alpha)$ in order to obtain rotating self-similar solutions to the scalar Helmholtz equation. Note, that we restrict our analysis to beams that are propagating in the positive $z$-direction.

For a beam that is self-accelerating three major properties need to be fulfilled. First, no external potential or non-linear optical effect should be present. Second, the beam is diffraction-free in a certain frame of reference. Finally, an observer resting in the aforementioned reference frame
would experience a fictitious force. The first condition is fulfilled immediately as we start our analysis from the linear and time-independent scalar Helmholtz equation. For the second requirement, a coordinate transformation needs to exist with which the field distribution is no longer dependent
on the propagation direction. It can easily be shown that an electric field of the general form
\begin{align}
E(r,\varphi,z)\overset{!}{=} E(r,\varphi+\omega z) \label{eq:condRot}
\end{align}
fulfils this condition. Obviously, with the substitution $\varphi^\prime=\varphi+\omega z$ the field $E(r,\varphi^\prime)$ in Eq.~\eqref{eq:condRot}
is no longer dependent on the longitudinal position $z$ and thus remains unchanged for every $z$. Moreover, the aforementioned coordinate
transformation describes a reference frame which is rotating with an angular velocity $\omega$. As a consequence, the last requirement is satisfied,
as an observer resting in this rotating frame of reference experiences a centrifugal force.

\newcommand*{\sign}{\mathrm{sign}}
Since Eq.~\eqref{eq:condRot} has to hold for every $\varphi$ and $z$, from Eq.~\eqref{eq:generalSuperpos} it immediately follows that
\begin{align}
\nicefrac{\beta}{n}\overset{!}{=}\omega. \label{eq:cond1}
\end{align}
As $\beta$ was restricted to be positive, the first conclusion from condition~\eqref{eq:cond1} is that the signs of $n$ and $\omega$ have to be
equal, i.e., $\sign(n)=\sign(\omega)$. Moreover, since the propagation constant $\beta$ is a function of the transverse component of the wave vector
$\alpha$, condition~\eqref{eq:cond1} can be rewritten as \begin{align} \alpha \overset{!}{=} \alpha_n = \sqrt{k^2-\omega^2n^2} \label{eq:cond2}
\end{align}
Obviously, this restriction can only be fulfilled for the specific choice of coefficients
\begin{align}
C_n(\alpha)=\widetilde{C}_n\,\delta(\alpha-\alpha_n).\label{eq:Cndelta}
\end{align}
Applying the restrictions on $\sign(n)$ as well as on $C_n(\alpha$), Eq.~\eqref{eq:generalSuperpos}  becomes
\begin{align}
E(r,\varphi,z)=\sum_{n=1}^{n_{\textrm{max}}} \widetilde{C}_n J_n(\alpha_n r)e^{i(\sign(\omega)n(\varphi+\omega z))}. \label{eq:Superpos}
\end{align}
This is the most general expression of a beam that rotates in a shape-invariant fashion with an angular velocity $\omega$. Note, that $n_{\textrm{max}}=\max\{n\in \mathbb{N}:\, k^2>\omega^2n^2 \}$ in order to ensure that evanescent waves are excluded from the sum. To give an intuitive description of this finding, it is helpful to consider the Fourier-transform of this  specific field. In essence, the Fourier-transform is a discrete superposition of concentric rings with radius $\alpha_n$ whereas the amplitude of these rings is given by the coefficients $\widetilde{C}_n$. Note that for a given $\omega$ Eq.~\eqref{eq:Cndelta} states that for each order $n$ there is exactly one ring radius $\alpha_n$. Moreover, each ring of order $n$ carries a helical phase pitch of $2\pi n$.

\begin{figure}
\centering
\includegraphics[width=\linewidth,page=5]{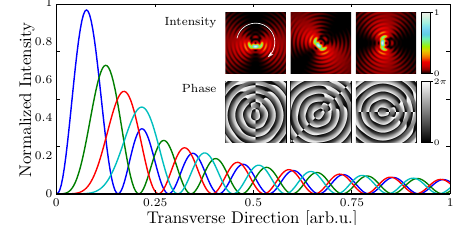}
\caption{Exemplary illustration of a radially self-accelerating field distribution with $\tilde{C}_{n}=1$ for $1\leq n\leq4$ and $\tilde{C}_{n}=0$
for $n>4$. The figure consists of  1-dimensional representation of the superimposed Bessel functions (main plot), resulting intensity distribution
(upper inset row) and resulting phase pattern (lower inset row).} \label{fig:example_field}
\end{figure}

%
%OLD VERSION
%As stated in the introduction,
%
The field pattern of a beam described by Eq.~\eqref{eq:Superpos} gives rise to screw-shaped trajectories, that show a non-degenerate periodicity in the azimuthal and the propagation direction. \autoref{fig:example_field} shows an appropriate example using four Bessel waves ranging from order $1$ to $4$, i.e., $\tilde{C}_{n}=1$ for $1\leq n \leq 4$ and $\tilde{C}_{n}=0$ for $n>4$. The depicted insets emphasize the fact that amplitude and phase are rotating synchronously as predicted. The angular frequency spectrum consists of four concentric rings with radii determined by Eq.~\eqref{eq:cond2}.

%\subsection{Rotating Intensity Distribution}
At the beginning of the theory section we indicated that for many practical applications, such as optical tweezing or micro-fabrication, only the intensity distribution of a beam might be of interest. For this reason, we want to state how the requirements for the beam profile change if condition \eqref{eq:condRot} is posed on the intensity, i.e., $I(r,\varphi,z)=\vert E(r,\varphi,z)\vert^2=I(r,\varphi+\omega z)$. It will be shown that this scenario is more general and offers a larger degree of freedom.  Consequently, in our subsequent discussion (including our experimental section) we will exclusively concentrate on this more general case. As the derivation of the following results is somewhat more lengthy and does not convey much additional physical insight, it is contained in the Supplementary Material.  However, from this derivation, one arrives at a constrain similar to \eqref{eq:cond2}, which reads
\begin{align}
\alpha_n=\sqrt{k^2-(\omega\,n+\beta_0)^2}. \label{eq:cond_I}
\end{align}
Moreover, under these conditions the field is given by
\begin{align}
E(r,\varphi,z)=e^{i\beta_0z}\sum_{n\in \mathcal{N}} \widetilde{C}_n J_n(\alpha_n r)e^{i(n(\varphi+\omega z))}. \label{eq:Superpos_I}
\end{align}
Note that there are two main differences between Eq.~\eqref{eq:Superpos_I} and Eq.~\eqref{eq:Superpos}. The first difference is that
Eq.~\eqref{eq:Superpos_I} contains the global phase factor $e^{i\beta_0z}$ with the propagation constant $\beta_0$, which can be regarded as a free
parameter for these beams. Consequently, one can already see that the field given by Eq.~\eqref{eq:Superpos_I} does not fulfill the requirement of a
rotation invariant field anymore (only the intensity is rotation invariant). It is important to be aware of the fact that $\beta_0$ does not only
determine the global phase factor but poses an important degree of freedom for scaling the transverse beam properties after fixing the rotation parameter $\omega$. This becomes apparent when comparing Eq.~\eqref{eq:cond2} and Eq.~\eqref{eq:cond_I}. The second difference is that the sum in Eq.~\eqref{eq:Superpos_I} contains also negative $n$, whereas Eq.~\eqref{eq:Superpos} covers only positive $n$. To be specific, the set $\mathcal{N}=\{n\in\mathbb{Z}:k^2>(\omega\,n+\beta_0)^2\}$ contains all integer numbers $n$ for which Eq.~\eqref{eq:cond_I} yields real values for
$\alpha_n$.

\begin{figure}
\centering
\includegraphics[width=\linewidth, page=4]{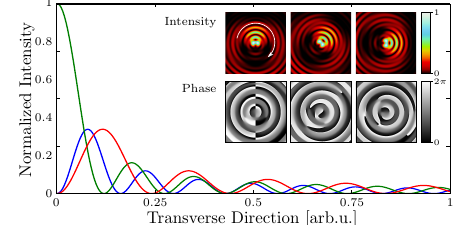}
\caption{Exemplary illustration of a radially self-accelerating intensity distribution with $\tilde{C}_{n}=1$ for $-1\leq n\leq +1$ and
$\tilde{C}_{n}=0$ for $\left|n\right|>1$. The figure consists of  1-dimensional representation of the superimposed Bessel functions (main plot), resulting
intensity distribution (upper inset row) and resulting phase pattern (lower inset row).} \label{fig:example_intensity}
\end{figure}

Please note that rotating intensity distributions based on pairs of superimposed Bessel-functions have been investigated theoretically and experimentally \cite{Pat96,Ter01, Vas09, Rop12, Rop12a, Ruf12,McG03,Paa98,Cha98}. Our theory, however, describes rotating field\,/\,intensity patterns that are composed of an arbitrary number of Bessel waves. The arising enlarged degree of freedom renders useful when modeling specific beam properties as will be discussed in the upcoming paragraph. \autoref{fig:example_intensity} shows an exemplary beam with $\tilde{C}_{n}=1$ for $-1\leq
n\leq +1$ and $\tilde{C}_{n}=0$ for $\left|n\right|>1$. The depicted insets demonstrate that the intensity distribution is indeed rotating during propagation while the corresponding phase is no longer synchronized. The angular frequency spectrum of the shown beam consists of three concentric
rings with radii that are determined by Eq.~\eqref{eq:cond_I}.

An important, yet open question is how versatile the transverse cross-section of beams described by Eq.~\eqref{eq:Superpos_I} can be tailored. In
order to answer this question consider Eq.~\eqref{eq:Superpos_I} in the initial plane ($z=0$).  For a given distance $R$ from the origin of the
coordinate system, Eq.~\eqref{eq:Superpos_I} can be written as
\begin{align}
E(R,\varphi,0)=\sum_{n\in\mathcal{N}} D_n e^{in\varphi}, \label{eq:Superpos_inital}
\end{align}
where $D_n=\widetilde{C}_n J_n(\alpha_n R)$.  If the distance $R$ is chosen such that $J_n(\alpha_n R)\neq 0$, then Eq.~\eqref{eq:Superpos_inital}
represents a general Fourier-series.  As an important consequence, the transverse beam profile can be tailored in a way that the field distribution
on a circle with radius $R$ can be chosen arbitrarily, in other words $E(R,\varphi,z=0)=f(\varphi)$, where the complex function $f(\varphi)$ can be
set without any condition.  From this it follows that a cooking recipe to tailor these kind of rotating beams could be to fix the parameter $\beta_0$
as well as the function $f(\varphi)$.  Then Eq.~\eqref{eq:cond_I} determines $\alpha_n$ and the expansion coefficients are given by
\begin{align}
\widetilde{C}_n =\frac{1}{2\pi J_n(\alpha_n R)}\int_{0}^{2\pi}f(\varphi)e^{-in\varphi}\mathrm{d}\varphi.
\end{align}
Note that as soon as the field is specified for one radius $R$, the entire field in the transverse plane is determined. This is simply due to the
fixed radial dependence of the Bessel functions.

%%%%%%%%%%%%%%%%%%%%%%%%%%%%%%%%%%%%%%%%%%%%%
%%%%%           Setup + Results         %%%%%
%%%%%%%%%%%%%%%%%%%%%%%%%%%%%%%%%%%%%%%%%%%%%

%\section{\label{sec:experiment}Experimental Setup and Results}
%\FloatBarrier

\begin{figure}
\includegraphics[width=\linewidth, page=2]{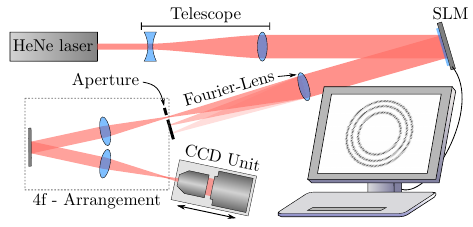}
\caption{Experimental setup containing a telescope for beam expansion, SLM (Holoeye Pluto VIS) for amplitude and phase modulation, lens for Fourier
transformation ($f=300\,\mathrm{mm}$), aperture and 4f-arrangement ($f_1=f_2=200\,\mathrm{mm}$) for signal cleaning as well as a movable CCD Unit (Basler
Ace1600-20gm with Olympus Plan N $20\times$) for data acquisition.} \label{fig:setup}
\end{figure}

In order to experimentally implement our findings, we make use of the fact that the presented beams show a multi-ring pattern with distinct helical
phase pitch in the angular frequency domain. Fourier transforming this pattern by means of a conventional lens will match the previously discussed
theory. For the experimental setup, different approaches are conceivable ranging from the use of axicons (conical lenses), ring slit apertures and
phase plates to the exclusive use of spatial-light-modulators (SLMs). In this work, we followed this last approach as it provides the highest amount
of flexibility. Our setup is presented in Fig.~\ref{fig:setup} and makes use of a technique introduced in Ref. \cite{Dav99}. This technique enables
simultaneous amplitude and phase modulation with a single phase-only SLM by multiplying the desired amplitude distribution with a blazed grating. In
our case this desired amplitude distribution is given by concentric rings - in other words, as previously mentioned, we implement the
Fourier-transform of the desired beam in the SLM-plane.  After the Fourier-transforming lens, undesired grating orders are filtered by a pin-hole and
the primary signal is imaged by an additional 4f-setup. Finally, a movable CCD-camera allows to measure the change of the intensity profile in
propagation direction.

With the proposed setup, we are able to cover almost the entire parameter range provided by our theory. This of course would go beyond the scope of this letter. We will therefore present an exemplary set of parameters upon which we show the practicability of radially self-accelerating beams and discuss the experimental limitations. One of these limitations is already given by the fact that Bessel beams cannot be created to their full extend as they would carry an infinite amount of energy and require a non-finite aperture. This is also true for the presented radially self-accelerating beams, since they are a specific discrete superposition of Bessel beams. Another limitation arises from the fact that the Fourier-transform of a Bessel wave is a ring with infinitesimal thickness. In an experimental setting it is clear that only rings with finite width
can be generated. As an immediate consequence, the range over which the experimentally generated beam resembles the theoretical prediction will be limited.

\begin{figure}
\centering
\includegraphics[width=\linewidth, page=3]{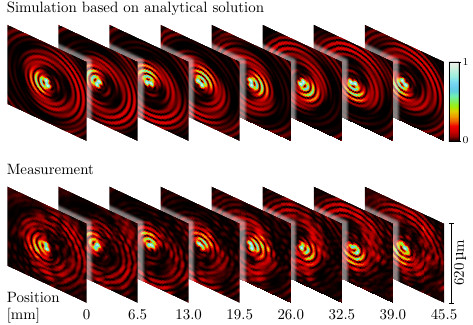}
\caption{Simulation (top) and experimental results (bottom) showing the propagation dynamics of a radially self-accelerating beam identical to the
one in Fig. \ref{fig:example_intensity}. Radii on the SLM have been $R_1=2.328\,\mathrm{mm}$, $R_2=1.958\,\mathrm{mm}$ and $R_3=1.501\,\mathrm{mm}$.}
\label{fig:measurement}
\end{figure}

%For experimental realization the set of parameters  Fig.~\ref{fig:example_intensity} 
For experimental realization the set of parameters presented in Fig. 3 was used.
For this purpose, a superposition of three rings was implemented in the SLM-plane. The subsequent Fourier-lens with focal length $f$ connects the ring radii $R_n$ on the SLM with the transverse components of the wave vector $\alpha_n$ via
\begin{align}
\alpha_n=\frac{k R_n}{f}.\label{equ:radii}
\end{align}
Fig.~\ref{fig:measurement} shows an experimental scan along the propagation direction together with a simulation based on the analytical solution
described by Eq.~\eqref{eq:Superpos_I}. In total, we were able to observe about two rotations over a length of $101.5\,\mathrm{mm}$. See Supplemental
Material at [URL will be inserted by publisher] for an animated representation of the full scan. The rotation rate was found to be
$\omega_{\mathrm{exp}}=\left(123.2\pm 2.4\right)\,\mathrm{\frac{rad}{m}}$ and is therefore in very good agreement with the intended value of
$\omega_{\mathrm{th}}=125\,\mathrm{\frac{rad}{m}}$.

%The slight deviation may be caused by alignment imperfections in the 6f-arrangement. As a result, the ring pattern prepared on the SLM would be scaled improperly thus violating Eq.~\eqref{eq:cond_I}.

%%%%%%%%%%%%%%%%%%%%%%%%%%%%%%%%%%%%%%%%%%%%%
%%%%%             Conclusion            %%%%%
%%%%%%%%%%%%%%%%%%%%%%%%%%%%%%%%%%%%%%%%%%%%%

%\section{\label{sec:conclusion}Conclusion}
In conclusion, we have demonstrated the existence of a new class of self-accelerating waves. Theoretically, those waves, which accelerate freely on
spiraling trajectories, were derived as solutions to the scalar Helmholtz equation. It was pointed out, that radially self-accelerating beams can be
generated as a discrete superposition of Bessel waves with well defined properties. As such, they are quasi-non-diffractive, meaning that they are
diffraction-free in a rotating, co-moving frame of reference. With the proposed experimental setup the study of beam properties under realistic
conditions was shown to be possible with great flexibility. In a first proof of principle experiment, it was verified that the beam shows indeed the
desired rotating behavior - yielding excellent agreement with the theoretical predictions - and, moreover, the transverse beam profile remains
unbroadened for a substantially long propagation distance.

%
%OLD VERSION
%As indicated in the introduction
%
We foresee a broad range of applications for this new class of radially accelerating beams ranging from particle
manipulation, e.g. as tractor beams, to material processing, e.g. photo lithography. In order to widen the range of possible applications even
further, it is also of interest to study the properties of these kind of beams in more detail on a fundamental basis. For instance, with particle
manipulation in mind, it is of great interest to investigate the self-healing behavior or the dynamics in random media. Regarding material
processing, for example, the dynamics of such field configurations in non-linear environments is of great importance.

% Further work will concern

%Presumably, this new class of self-accelerating beams will inspire as many applications as previous self-accelerating beams did. In particular we expect them to be useful for particle manipulation (e.g. tractor beams) and material processing (e.g. photo lithography).

%%%%%%%%%%%%%%%%%%%%%%%%%%%%%%%%%%%%%%%%%%%%%
%%%%%           Acknowledgments         %%%%%
%%%%%%%%%%%%%%%%%%%%%%%%%%%%%%%%%%%%%%%%%%%%%

%\FloatBarrier
\begin{acknowledgments}
The authors wish to thank the German Ministry of Education and Research (Center for Innovation Competence program, grant 03Z1HN31), the Thuringian Ministry for Education, Science and Culture (Research group Spacetime, grant no. 11027-514), the Deutsche Forschungsgemeinschaft (grant NO462/6- 1), and the German-Israeli Foundation for Scientific Research and Development (grant 1157-127.14/2011).
\end{acknowledgments}

%%%%%%%%%%%%%%%%%%%%%%%%%%%%%%%%%%%%%%%%%%%%%
%%%%%              Appendix             %%%%%
%%%%%%%%%%%%%%%%%%%%%%%%%%%%%%%%%%%%%%%%%%%%%

%\appendix
%\section{Further Information} starred version if just one section is used
%\section{Even further information}

%%%%%%%%%%%%%%%%%%%%%%%%%%%%%%%%%%%%%%%%%%%%%
%%%%%           Bibliography            %%%%%
%%%%%%%%%%%%%%%%%%%%%%%%%%%%%%%%%%%%%%%%%%%%%

\FloatBarrier
\bibliography{myliterature}{}

\end{document}